\documentclass[preprint,superscriptaddress,aps]{revtex4}
\usepackage[latin9]{inputenc}
\usepackage{amsfonts}
\usepackage{graphics}
\usepackage{graphicx}
\usepackage{slashed}
\usepackage{indentfirst}            % Recua o primeiro parágrafo de cada seção
\usepackage{microtype}              % Melhora a justificação do documento
\usepackage{subeqnarray}            % Subenumeração de equações
\usepackage{amsmath}                % Funções matemáticas
\usepackage{mathtools}
\usepackage{calrsfs}
\begin{document}
	
	\newcommand{\be}{\begin{equation}}
		\newcommand{\ee}{\end{equation}}
	\newcommand{\en}{\end{equation}}
\newcommand{\ba}{\begin{eqnarray}}
	\newcommand{\ea}{\end{eqnarray}}
\newcommand{\bea}{\begin{eqnarray}}
	\newcommand{\eea}{\end{eqnarray}}
\newcommand{\bq}{\begin{eqnarray}}
	\newcommand{\eq}{\end{eqnarray}}
\newcommand{\dt}{\frac{d^3k}{(2 \pi)^3}}
\newcommand{\dtp}{\frac{d^3p}{(2 \pi)^3}}
\newcommand{\kbruto}{\hbox{$k \!\!\!{\slash}$}}
\newcommand{\pbruto}{\hbox{$p \!\!\!{\slash}$}}
\newcommand{\qbruto}{\hbox{$q \!\!\!{\slash}$}}
\newcommand{\lbruto}{\hbox{$l \!\!\!{\slash}$}}
\newcommand{\bbruto}{\hbox{$b \!\!\!{\slash}$}}
\newcommand{\parbruto}{\hbox{$\partial \!\!\!{\slash}$}}
\newcommand{\Abruto}{\hbox{$A \!\!\!{\slash}$}}
\newcommand{\pa}{\partial}
\def\pls{\partial\!\!\!/}
\def\bb{\bibitem}
\def\as{a\!\!\!/}
\def\As{A\!\!\!/}
\def\ks{k\!\!\!/}
\def\ls{l\!\!\!/}
\def\ps{p\!\!\!/}
\def\qs{q\!\!\!/}
\def\bs{b\!\!\!/}
\def\yb{\bar{\y}}
\def\Ds{D\!\!\!\!/}
\def\ds{\partial\!\!\!/}
\newcommand{\fr}{\frac}
\def\ov{\over}
\def\g{\gamma}
\def\n{\nu}
\def\m{\mu}
\def\n{\nu}
\def\eff{\mathrm{eff}}
\def\CS{\mathrm{CS}}
\newcommand{\Slash}[1]{{#1}\!\!/}
\newcommand{\SLASH}[1]{{#1}\!\!\!/}
\newcommand{\RM}[1]{\mathrm{#1}}
%\bibliographystyle{apsrev4-1}  % ou 'apsrev', 'plain', etc., dependendo do estilo desejado
%\bibliography{referencias}     % nome do seu arquivo .bib (sem extensão)

\title{Classical investigations in a CPT-even Lorentz-violating model and their implications for the Compton effect}

\date{today}

\author{E. Neres J\'unior} \email[]{edson.neres@ufvjm.edu.br}
\affiliation{Centro Federal de Educa\c{c}\~ao Tecnol\'ogica - MG, Avenida Amazonas, 7675 - 30510-000 - Nova Gameleira - Belo Horizonte-MG, Brasil}
\affiliation{Instituto de Engenharia, Ci\^encia e Tecnologia, Universidade Federal dos Vales do Jequitinhonha e Mucuri, Avenida Um, 4050 - 39447-790 -Cidade Universit\'aria -Jana\'uba-MG, Brasil}

\author{J. C. C. Felipe} \email[]{jeanccfelipe@ufsj.edu.br}
\affiliation{Departamento de Estat\' istica, F\'isica e Matem\'atica, Universidade Federal de S\~{a}o Jo\~{a}o del Rei, Rod. MG 443, Km 7, 36497-899 - Ouro Branco-MG, Brasil}

\author{A. P. Ba\^eta Scarpelli} \email[]{scarpelli@cefetmg.br}
\affiliation{Centro Federal de Educa\c{c}\~ao Tecnol\'ogica - MG, Avenida Amazonas, 7675 - 30510-000 - Nova Gameleira - Belo Horizonte-MG, Brasil}

\author{A. Yu. Petrov}
\email[]{petrov@fisica.ufpb.br}
\affiliation{Departamento de F\'{\i}sica, Universidade Federal da Para\'{\i}ba,\\
 Caixa Postal 5008, 58051-970, Jo\~ao Pessoa, Para\'{\i}ba, Brazil}
 
\author{J. A. Helay\"{e}l-Neto}
\email[]{helayel@cbpf.br}
\affiliation{Centro Brasileiro de Pesquisas F\'{\i}sicas, \\Rua Dr. Xavier Sigaud 150, Urca, RJ 22290-180, Brazil}

\begin{abstract}
	In this work, we investigate some aspects of the Maxwell electrodynamics with the additive Lorentz-violating (LV) CPT-even term. For this model, we derive the energy and momentum conservation laws, highlighting the modifications introduced by Lorentz violation. Furthermore, using the modified dispersion relations, we analyze the correction to Compton effect arising from the presence of the LV vector.
\end{abstract}

\maketitle

\section{Introduction}
In the studies of quantum field theory, considering models with violation of Lorentz symmetry arises from different approaches that seek to go beyond the Standard Model of particle physics. As it is well known, Lorentz symmetry is one of the fundamental bases of Special Relativity and, by extension, of all relativistic field theories. However, several theories based on a natural need to unify gravitation with quantum mechanics, such as quantum gravity and string theory, suggest that this symmetry may only be a valid approximation on relatively low energy scales. It is natural to expect that at Planck scales, where quantum gravitational effects become significant, Lorentz symmetry could be violated or modified. In this context, models that contemplate the violation of this symmetry provide a framework for investigating phenomena that could be detected in high-precision experiments, such as possible deviations from the Standard Model (SM) predictions within fundamental interactions observed in particle accelerators or through high-energy astrophysical measurements \cite{Kosteleck_Samuel}.

In addition to theoretical predictions, the experimental search for signs of violation of Lorentz symmetry is an important motivation for studying these models as well. Since the 1990s, several experiments have been designed with the aim of testing the limits of this symmetry with very high precision. For example, experiments with atomic clocks, astrophysical observations of ultra-high-energy cosmic rays, and studies of particle scattering at accelerators such as the LHC have placed increasingly tighter constraints on possible signs of Lorentz symmetry violation \cite{Mattingly}. A paradigmatic review on most recent experimental results on LV effects with estimations for many LV parameters is presented in \cite{Kostelecky:2008ts}. These investigations are motivated by the possibility that, by detecting a small violation, new possibilities to physics beyond the Standard Model will be appear, allowing predictions of theories of quantum gravity or other extensions of current physics to be tested.

The CPT-even terms have attracted significant interest, partly due to their conceptual similarity to the notion of the aether in pre-relativistic theories and their role in scenarios involving extra dimensions (see the discussion in \cite{Carroll}). In many cases, Lorentz symmetry violation is associated with the presence of a preferred spacetime background presented by some constant vector, or, in general, a tensor. It has been suggested, for example, that Lorentz-violating tensor fields with vacuum expectation values aligned along extra dimensions can effectively hide these dimensions from low-energy physics \cite{Carroll}. In such frameworks, Lorentz-violating terms in the gauge, fermionic, and scalar sectors may originate from interactions with background aether-like fields. The CPT-even term considered in this work can be radiatively induced when a nonminimal Lorentz-violating coupling is introduced into electrodynamics \cite{Petrov,Scarp,Scarp2,Petrov-Scarp}.

Within this scenario, among various studies of scattering processes in the presence of LV effects, see e.g. \cite{Charneski:2012py,Ferreira:2004hx,Casana:2012vu,dosSantos:2018rns}, the Compton effect displays an interesting scenario when LV effects are included \cite{Cabral2023, Aultschul2004, Brito2016}. It is interesting to observe that we have an increasing photon wavelength after scattering when we consider these effects, which is important to be analyzed, given that various experimental data were obtained through this effect \cite{Feldman2012, Myers2014} as well as through its nonlinear generalization \cite{Hela5}.

Within this study, the contribution relevant for us arises from the addition of a CPT-even LV added to the Maxwell Lagrangian. The term we are interested in is a particular case of the one present in the Minimal Standard Model Extension (MSME), defined in \cite{SME1,SME2}. The MSME includes, in the minimal Standard Model Lagrangian density, all possible terms that may emerge as a consequence of spontaneous Lorentz symmetry breaking at very high energies.
These terms involve constant tensors that can lead to unconventional effects, such as spacetime anisotropy \cite{belich, APetrov}. In the scenario analyzed in the present work, however, we consider the case where the background vector depends explicitly on spacetime coordinates.

This article is organized as follows. In Section \ref{sec1}, we present Maxwell equations modified by the inclusion of the symmetry-violating term in the Lagrangian density. In the Subsection \ref{Ener_violadas}, we obtain the generalized expressions for the Lorentz force density, momentum density, Poynting vector, and energy density, while in the Subsection \ref{Campo_eletrico}, we analyze the electric field generated by a point charge in a Lorentz symmetry-violating background. In Section \ref{Ef_compton}, we investigate the dispersion relations and the kinematics of the Compton effect within this context. Finally, in Section IV, we summarize our results.

\section{The LV electrodynamics}\label{sec1}
We begin by considering a CPT-even LV electrodynamics described by the following Lagrangian density,
\be
\label{ourl}
{\cal L}= -\frac{1}{4 \mu_0}F_{\mu\nu}F^{\mu\nu} 
-\frac{1}{2 \mu_0}(\chi_\mu F^{\mu\nu})^2 - J_\mu A^\mu, 
\ee
where $\chi^\mu(x)$ is a $x^\mu$ dependent background vector, $\mu_0$ is the magnetic constant, $J^\mu$ is the conserved current, and we use the SI units, to simplify the comparison with standard electrodynamics textbooks. The term $-\frac{1}{2}\left(\chi_\mu F^{\mu \nu}\right)^2$ is a special case of the CPT-even one in the photon sector of the minimal SME, which is given by 
\begin{equation}\label{tensor_kapa}
	{\cal L}_{\text {even}}=-\frac{1}{4} \kappa^{\alpha \beta \gamma \delta} F_{\alpha \beta} F_{\gamma \delta},
\end{equation}
where $\kappa^{\alpha\beta\lambda\delta}$ is a  fourth-rank tensor \cite{PhysRevD.69.105009}. Its particular form presented in (\ref{ourl}), also referred as the aether term \cite{Petrov,Petrov-Scarp}, is reproduced for the particular choice
%na forma 
%\begin{equation}
%\mathscr{L}_{\text {par }}=\chi^\alpha \chi_\gamma F_{\alpha \beta} F^{\gamma \beta},
%\end{equation}
\begin{equation}\label{ex_tenso_rank4}
	\kappa_{\alpha \beta \gamma \delta}=-\left(\chi_\alpha \chi_\gamma \eta_{\beta \delta}-\chi_\alpha \chi_\delta \eta_{\beta \gamma}+\chi_\beta \chi_\delta \eta_{\alpha \gamma}-\chi_\beta \chi_\gamma \eta_{\alpha \delta}\right).
\end{equation}
The modified field equations, in this case, are given by
\be
\partial_\nu \left[F^{\mu\nu} - \chi_\alpha \chi^\nu F^{\alpha \mu} + \chi_\alpha \chi^\mu F^{\alpha \nu}    \right] = -\mu_0 J^\mu,
\label{fequations}
\ee
and, besides, the standard Bianchi identity is maintained:
\be
\partial_\nu \Tilde{F}^{\mu\nu}=0.
\ee
In order to continue our studies, we first derive the modified Gauss equation by taking $\nu=0$ in equation (\ref{fequations}). We obtain
\be
\boldsymbol{\nabla}\cdot \boldsymbol{D}=\frac{\rho}{\varepsilon_0}. \label{Gauss}
\ee
The electric displacement is now modified, having the form
\be
\boldsymbol{D}=(1+ \chi_0^2)\boldsymbol{E} + c \chi_0 (\boldsymbol{\chi} \times \boldsymbol{B}) - (\boldsymbol{\chi} \cdot \boldsymbol{E})\boldsymbol{\chi}.
\ee
For the Amp\`ere-Maxwell equation, we take the spatial index $i$ in equation (\ref{fequations}), so that
\be
\boldsymbol{\nabla} \times \boldsymbol{H} = \mu_0 \varepsilon_0\frac{\partial \boldsymbol{D}}{\partial t} + \mu_0 \boldsymbol{J}, \label{Ampere}
\ee
where
\be
\boldsymbol{H}=\boldsymbol{B} + \frac 1c \chi_0 (\boldsymbol{\chi} \times \boldsymbol{E}) +\boldsymbol{\chi} \times (\boldsymbol{\chi} \times \boldsymbol{B}).
\ee
From the Bianchi identity, the two usual homogeneous equations are written as in the standard case:
\be
\boldsymbol{\nabla}\times \boldsymbol{E} = - \frac{\partial \boldsymbol{B}}{\partial t}
\;\;\;\; \mbox{and} \;\;\;\; \boldsymbol{\nabla} \cdot \boldsymbol{B} = 0. \label{Faraday}
\ee

Now, as we have the modified Maxwell equations, it is interesting to investigate what kind of modifications it would be possible to observe in physical processes. In the following subsections, we carry out some of these analyses.

\subsection{The modification of the energy conservation law}\label{Ener_violadas}

The first question we would like to discuss is how the conservation of momentum and energy is affected by the inclusion of this kind of LV term.

First, we obtain the scalar product of the Faraday-Lenz equation with $\boldsymbol{B}$. We get
\begin{equation}\label{Far_lenz_B}
	\left(\boldsymbol{\nabla}\times\boldsymbol{E}\right)\cdot\boldsymbol{B}=-\frac{1}{2}\partial_t\left(\boldsymbol{B}^2\right).
\end{equation}
We also find the scalar product of the Amp\`{e}re-Maxwell equation with $\boldsymbol{E}$ which yields
\begin{eqnarray}\label{Amp_Max_E}
	&\boldsymbol{\nabla}\times\left(\boldsymbol{B}+\boldsymbol{\chi}\times\left(\boldsymbol{\chi}\times\boldsymbol{B}\right)+\frac{1}{c}\chi_0\boldsymbol{\chi}\times\boldsymbol{E}\right)\cdot\boldsymbol{E}=\mu_0\boldsymbol{J}\cdot\boldsymbol{E}+&\nonumber\\
	&+\mu_0\varepsilon_0\partial_t\left(\left(1+\chi_{_0}^2\right)\ \boldsymbol{E}+c\chi_{_0}\left(\boldsymbol{\chi} \times \boldsymbol{B}\right) -\boldsymbol{\chi} \left(\boldsymbol{\chi} \cdot \boldsymbol{E}\right)\right)\cdot\boldsymbol{E}.&
\end{eqnarray}
Now we take the difference between equations (\ref{Amp_Max_E}) and (\ref{Far_lenz_B}), so that we get
\begin{eqnarray}\label{eq_Enr_1}
	&\boldsymbol{\nabla}\cdot\left(\boldsymbol{E}\times\boldsymbol{B}\right)-\boldsymbol{\nabla}\times\left(\boldsymbol{\chi}\times\left(\boldsymbol{\chi}\times\boldsymbol{B}\right)+\frac{1}{c}\chi_0\boldsymbol{\chi}\times\boldsymbol{E}\right)\cdot\boldsymbol{E}=-\frac{1}{2}\partial_t\left(\boldsymbol{B}^2\right)+&\nonumber\\ &- \mu_0\boldsymbol{J}\cdot\boldsymbol{E} - \mu_0\varepsilon_0\partial_t\left(\left(1+\chi_{_0}^2\right)\ \boldsymbol{E}+c\chi_{_0}\left(\boldsymbol{\chi} \times \boldsymbol{B}\right) -\boldsymbol{\chi} \left(\boldsymbol{\chi} \cdot \boldsymbol{E}\right)\right)\cdot\boldsymbol{E}.&
\end{eqnarray}
After a long sequence of algebraic manipulations, we finally reach the continuity-like equation 
\begin{eqnarray}\label{eq_continuidade_4}
	\frac{\partial \widetilde{U}}{\partial t}+\boldsymbol{\nabla} \cdot \boldsymbol{\widetilde{S}}&=&-\boldsymbol{E} \cdot \boldsymbol{J}-\frac{1}{2 \mu_0} \boldsymbol{B}^2 \partial_t\left(\boldsymbol{\chi}^2\right)+\frac{1}{\mu_0} \boldsymbol{\chi} \cdot \boldsymbol{B} \boldsymbol{B} \cdot {\partial}_t(\boldsymbol{\chi}) + \nonumber\\
	& &-\frac{\varepsilon_0}{2} \boldsymbol{E}^2 \partial_t\left(\boldsymbol{\chi}_0^2\right)+c \varepsilon_0(\boldsymbol{E} \times \boldsymbol{B}) \cdot \partial_t\left(\chi_0 \boldsymbol{\chi}\right) + \nonumber\\
	& &+ \varepsilon_0 \boldsymbol{\chi} \cdot \boldsymbol{E} \boldsymbol{E} \cdot \partial_t(\boldsymbol{\chi}),
\end{eqnarray}
in which the modified energy density and Poynting vector are defined, respectively, as
\begin{equation}
	\widetilde{U}\equiv\frac{1}{2}\left(\epsilon_0\boldsymbol{E}^2+\frac{1}{\mu_0} \boldsymbol{B}^2 + \frac{1}{\mu_0}(\boldsymbol{\chi} \cdot \boldsymbol{B})^2 -\frac{1}{\mu_0} \boldsymbol{\chi}^2 \boldsymbol{B}^2  + \varepsilon_0 \chi_0^2\boldsymbol{E}^2 -\varepsilon_0(\boldsymbol{\chi} \cdot \boldsymbol{E})^2 \right)
\end{equation}
and
\begin{equation}
	\boldsymbol{\widetilde{S}}\equiv\frac{1}{\mu_0}(\boldsymbol{E} \times \boldsymbol{B}) +\frac{1}{\mu_0}(\boldsymbol{E} \times \boldsymbol{\chi}) \boldsymbol{\chi} \cdot \boldsymbol{B} +\frac{1}{\mu_0}(\boldsymbol{B} \times \boldsymbol{E}) \boldsymbol{\chi}^2 +c \varepsilon_0 \chi_0 \boldsymbol{E} \times(\boldsymbol{\chi} \times \boldsymbol{E}).
\end{equation}
It is easy to observe that the results of Maxwell electrodynamics are recovered if the background vector $\chi^{\mu}$ is chosen to be zero.

In the same sense, we perform the vectorial products of the Faraday-Lenz equation with the field $\boldsymbol{E}$ and of the 
Amp\`{e}re-Maxwell equation with the field $\boldsymbol{B}$. Then, the two resulting equations are combined to obtain a new continuity equation, which incorporates the momentum and force densities, corrected by the violation terms:
\begin{equation}\label{forca de lorentz violada}
	\frac{1}{c^2} \frac{\partial \mathbf{\widetilde{g}}}{\partial t}+\nabla \cdot \overleftrightarrow{\mathbf{\widetilde{T}}}=-\rho \mathbf{E}-\mathbf{J} \times \mathbf{B} + \Theta,
\end{equation}
where $\mathbf{\widetilde{g}}$ represents the momentum density, $\overleftrightarrow{\mathbf{\widetilde{T}}}$ is the tension tensor of Maxwell, and  $\Theta$ corresponds to the correction to the force density. These objects are given by 
\begin{eqnarray}
	\mathbf{\widetilde{g}}= \frac{1}{\mu_0}\left\{\boldsymbol{E} \times \boldsymbol{B}+ \chi_0^2 \boldsymbol{E} \times \boldsymbol{B}+c \chi_0(\boldsymbol{\chi} \times \boldsymbol{B}) \times \boldsymbol{B}-(\boldsymbol{\chi} \times \boldsymbol{B})(\boldsymbol{\chi} \cdot \boldsymbol{E})\right\},
\end{eqnarray}
\begin{eqnarray}
	\Theta&=& \frac{1}{2\mu_0}\boldsymbol{B}^2\boldsymbol{\nabla}\boldsymbol{\chi}^2-\frac{1}{\mu_0}(\boldsymbol{\chi}\cdot\boldsymbol{B})(\boldsymbol{B}\cdot\partial_i\boldsymbol{\chi}) + c \varepsilon_0(\boldsymbol{B} \times \boldsymbol{E})_j \partial_i\left(\chi_0 \chi_j\right)\nonumber\\
	& & + \frac{1}{2} \varepsilon_0 \boldsymbol{E}^2 \partial_i\left(\chi_0^2\right) -\varepsilon_0 (\boldsymbol{\chi} \cdot \boldsymbol{E}) (\boldsymbol{E} \cdot \partial_i(\chi))
\end{eqnarray}
and
\begin{eqnarray}
	&(\overleftrightarrow{\mathbf{\widetilde{T}}})_{i j}=\partial_j\left\{\delta_{ij}\left(\frac{\varepsilon_0}{2} \boldsymbol{E}^2+\frac{1}{2 \mu_0} \boldsymbol{B}^2-\frac{1}{2 \mu_0}(\boldsymbol{\chi} \times \boldsymbol{B})^2+\frac{\varepsilon_0}{2} \boldsymbol{E}^2 \chi_0^2-\frac{\varepsilon_0}{2} \boldsymbol{\chi} \cdot \boldsymbol{E}\right) +\right.&\nonumber\\
	& \left.-\varepsilon_0 E_i E_j-\frac{1}{\mu_0} B_i B_j-\frac{1}{\mu_0}(\boldsymbol{\chi} \times(\boldsymbol{\chi} \times \boldsymbol{B}))_i B_j-c \varepsilon_0 \chi_0(\boldsymbol{\chi} \times \boldsymbol{E})_i B_j+\right.&\nonumber\\
	&\left.-\varepsilon_0 E_i E_j\chi_0^2 - c \varepsilon_0 \chi_0(\boldsymbol{\chi} \times \boldsymbol{B})_j E_i + \varepsilon_0(\boldsymbol{\chi} \cdot \boldsymbol{E}) \chi_j E_i \right\}.&
\end{eqnarray}
Again, it is easy to check that in the limit $\chi^{\mu}=0$, in which the Lorentz symmetry is restored, the standard Maxwell equations are reproduced.

\subsection{Electric field for a point charge}\label{Campo_eletrico}

Considering a point particle located at the origin and carrying an electric charge q, in Gaussian units, we obtain the following equations from Eqs. (\ref{Gauss}), (\ref{Ampere}), and (\ref{Faraday}):

\begin{eqnarray}
	\boldsymbol{\nabla} \cdot\left[(1+\chi_0^2) \boldsymbol{E}+c \chi_0\left(\boldsymbol{\chi} \times \boldsymbol{B}\right)-\boldsymbol{\chi}(\boldsymbol{\chi} \cdot \boldsymbol{E})\right] &=& \frac{1}{\epsilon_0} q \delta^3(\boldsymbol{r}),\label{gaus_carga_pontual}\\
	\boldsymbol{\nabla} \times \boldsymbol{E}&=&\boldsymbol{0},\label{rotE_nulo}\\  \boldsymbol{\nabla} \cdot \boldsymbol{B}&=&0,\\
	\boldsymbol{\nabla} \times\left(\boldsymbol{B}+\boldsymbol{\chi} \times(\boldsymbol{\chi} \times \boldsymbol{B})+\frac{1}{c} \chi_0 \boldsymbol{\chi} \times \boldsymbol{E}\right)&=&\boldsymbol{0}.\label{eq_amper_carga_pont}
\end{eqnarray}
where $\epsilon_0$ is the vacuum electric permittivity. It is worth noting that the vacuum can exhibit nontrivial dielectric permittivity and magnetic permeability in scenarios involving Lorentz symmetry breaking \cite{Hela1}. This particular feature motivates further investigations in future work. Given that we are dealing with a point charge located at the origin, and recalling that the curl of the gradient of any continuous scalar field $\alpha$ is identically zero, we obtain from equation (\ref{eq_amper_carga_pont}) the following result
\begin{eqnarray}\label{B_fun_de_E}
	\boldsymbol{B}+\boldsymbol{\chi} \times(\boldsymbol{\chi} \times \boldsymbol{B})+\frac{1}{c} \chi_0 \boldsymbol{\chi} \times \boldsymbol{E}=\boldsymbol{\nabla}\alpha.
\end{eqnarray}
It is clear that in principle, without loss of generality, the scalar field $\alpha$ can be assumed to vanish. Accordingly, we can rewrite the equation (\ref{B_fun_de_E}) as follows:
\begin{eqnarray}
	M_{ij}B_j=P_i,
\end{eqnarray}
with
\begin{eqnarray}
	M_{ij}=\left[1-\boldsymbol{\chi}^2\right] \delta_{i j}+\chi_i \chi_j,
\end{eqnarray}
and
\begin{equation}
	P_i=\left(\boldsymbol{\nabla}\alpha -\frac{\chi_0}{c}\boldsymbol{\chi} \times \boldsymbol{E}\right)_i.    
\end{equation}
Since
\begin{eqnarray}
	M_{j m}^{-1}=\frac{1}{1-\boldsymbol{\chi}^2}\left\{\delta_{jm} - \chi_j \chi_{m}\right\},
\end{eqnarray}
we obtain $B$ from $M^{-1}_{jm}P_m$. Therefore, we have
\begin{eqnarray}\label{valor_B}
	\boldsymbol{B}=\frac{1}{1-\boldsymbol{\chi}^2}\left\{\boldsymbol{\nabla}\alpha-\boldsymbol{\chi}(\boldsymbol{\chi}\cdot\boldsymbol{\nabla}\alpha)\right\}-\frac{\chi_0}{c(1-\boldsymbol{\chi}^2)}\left(\boldsymbol{\chi}\times\boldsymbol{E}\right).
\end{eqnarray}
We thus observe a remarkable effect of Lorentz symmetry violation: the emergence of a magnetic field in the presence of a static, point-like electric charge. This phenomenon arises from the interaction between the electric field generated by the charge and the background field implementing the Lorentz symmetry breaking. Such an effect does not occur in conventional electrodynamics, where a static charge does not produce a magnetic field. 

On the other hand, from equation (\ref{gaus_carga_pontual}) and the result obtained in (\ref{valor_B}) we have

\begin{eqnarray}\label{eq_para_E}
	\boldsymbol{E}-(\boldsymbol{\chi}\cdot\boldsymbol{E})\boldsymbol{\chi}=\frac{1}{f}\left(\boldsymbol{\xi}-\boldsymbol{\theta}\right),
\end{eqnarray}
with
\begin{eqnarray}\label{def_f}
	f=\frac{1+\chi_0^2-\boldsymbol{\chi}^2}{1-\boldsymbol{\chi}^2},\; \boldsymbol{\theta}=\frac{c\chi_0}{1-\boldsymbol{\chi}^2}\left(\boldsymbol{\chi}\times\boldsymbol{\nabla}\alpha\right)\;\mbox{e} \;\boldsymbol{\xi}=\frac{q}{4 \pi \epsilon_0 r^2} \hat{r}.
\end{eqnarray}

In a similar way to that one followed for the above study of the magnetic field, we now write the equation for the electric field (\ref{eq_para_E})  as follows 
\begin{eqnarray}
	M_{ij}E_j=\frac{1}{f}(\xi-\theta)_i,
\end{eqnarray}
with
\begin{eqnarray}
	M_{ij}=\delta_{ij}-\chi_i\chi_j.
\end{eqnarray}
whose inverse is given by
\begin{eqnarray}
	M_{j m}^{-1}= \delta_{j m}+\frac{1}{\left(1-\boldsymbol{\chi}^2\right)} \chi_j \chi_m.
\end{eqnarray}
Therefore, the electric field for a point charge in the origin of the coordinate system is given by 
\begin{eqnarray}\label{campoE}
	\boldsymbol{E}=\frac{1}{f} \frac{q}{4 \pi \epsilon_0 r^2} \hat{r}-\frac{c\chi_0}{1-\boldsymbol{\chi}^2+\chi_0^2}\boldsymbol{\chi}\times\boldsymbol{\nabla}\alpha+\frac{q}{\left(1-\boldsymbol{\chi}^2+\chi_0^2\right) 4 \pi \epsilon_0 r^2} \boldsymbol{\chi}(\boldsymbol{\chi} \cdot \hat{r}),
\end{eqnarray}
where $\boldsymbol{r}=(x, y, z)$ is the position vector of the point where the field is calculated, $r=|\boldsymbol{r}|=\sqrt{x^2+y^2+z^2}$ is the distance of the charge to that point and $\hat{r}=\frac{\boldsymbol{r}}{r}$. Here, a comment is in order. Note that the electric field $\boldsymbol{E}$, in this formulation, is not purely radial, presenting a deviation from the central field. This contrasts with conventional electrodynamics, in which the field points exclusively in the radial direction.

However, let us note that the electric field $\boldsymbol{E}$ found in (\ref{campoE}) must satisfy the equation (\ref{rotE_nulo}). This implies that the field $\boldsymbol{E}$ must be derived from a single scalar potential $\phi$, whereas $\boldsymbol{\nabla}\times(-\boldsymbol{\nabla}\phi)=\boldsymbol{0}$. Taking into account the second term, we conclude that either $\boldsymbol{\chi}$ is parallel to $\boldsymbol{\nabla}\alpha$, or $\boldsymbol{\nabla}\alpha=\boldsymbol{0}$, so that the curl of this term is zero. Considering $\boldsymbol{\chi}=g_1(r)\hat{r}$ and $\chi_0=g_2(r)$ in
(\ref{campoE}), we can assume, without loss of generality, that $\boldsymbol{\nabla}\alpha=\boldsymbol{0}$, or otherwise $\boldsymbol{\nabla\alpha}$ would also have the direction $\hat{r}$. Thus, the electric field can be rewritten from the equation (\ref{campoE}) in the form
\begin{eqnarray}\label{campo_E2}
	\boldsymbol{E} &=& h(r)\frac{q}{4\pi\epsilon_0r^2}\hat{r},
\end{eqnarray}
where $h(r)=f_1(r)+f_2(r)g_1^2(r)$, with $f_1(r)=\frac{1}{f}$, $f_2(r)=\frac{1}{1-\chi^2+\chi_0^2}$. This electric field satisfies the relation $\boldsymbol{\nabla}\times\boldsymbol{E}=\boldsymbol{0}$. 
Note that the electric field is modified by the violation parameter. Another relevant point is that, in the limit $\chi^\mu \rightarrow 0$, which leads to $h \rightarrow 1$ in (\ref{campo_E2}), we were able to recover the corresponding expression for the Maxwell Electrodynamics. 

\section{Dispersion relations and the kinematics of the Compton effect}\label{Ef_compton}

In this section, we derive the dispersion relations for the particular case in which the LV parameter is time-independent, and purely spacelike, $\chi^\mu=(0, \boldsymbol{\chi})$. Besides, we consider the situation in the absence of sources $(\vec{J}=0$ and $\rho=0)$. With these conditions, starting from the equations (\ref{Gauss}), (\ref{Ampere}) and (\ref{Faraday}), we have
\begin{eqnarray}
	\boldsymbol{\nabla} \cdot\left( \boldsymbol{E}-\boldsymbol{\chi}(\boldsymbol{\chi} \cdot \boldsymbol{E})\right) &=& 0,\label{gaus_mod}\\
	\boldsymbol{\nabla} \times \boldsymbol{E}=-\partial_t\boldsymbol{B},& & \boldsymbol{\nabla} \cdot \boldsymbol{B}=0, \label{homog}\\  
	\boldsymbol{\nabla} \times\left(\boldsymbol{B}+\boldsymbol{\chi} \times(\boldsymbol{\chi} \times \boldsymbol{B})\right)&=&\mu_0\epsilon_0\partial_t\left( \boldsymbol{E}-\boldsymbol{\chi}(\boldsymbol{\chi} \cdot \boldsymbol{E})\right).\label{ampere_modificada}
\end{eqnarray}

Considering the solutions for the plane wave $\boldsymbol{E}(\boldsymbol{x}, t)=\boldsymbol{E}_0 e^{i(\boldsymbol{k} \cdot \boldsymbol{x}-\omega t)}$ and $\boldsymbol{B}(\boldsymbol{x}, t)=\boldsymbol{B}_0 e^{i(\boldsymbol{k} \cdot \boldsymbol{x}-\omega t)}$ in equations (\ref{gaus_mod}), (\ref{homog}) and (\ref{ampere_modificada}), with $\boldsymbol{E}_0$ and $\boldsymbol{B}_0$ are constant vectors, the relationship between frequency $\omega$ and the wave vector $\boldsymbol{k}$ could be written in the matrix form, as we can see below
\begin{equation}\label{eq_matricial}
	M_{i j} E_{0 j}=0,
\end{equation}
where $E_{0 j}$ $(j=1, 2, 3)$ are the electric field components. The matrix $M_{i j}$ has the form
\begin{equation}
	M_{i j}=\alpha \delta_{i j}+\beta u_i v_j,
\end{equation}
in which
\begin{eqnarray}
	& & \alpha= \mu_0\epsilon_0\omega^2-\boldsymbol{k}^2+(\boldsymbol{\chi} \cdot \boldsymbol{k})^2,\quad\beta=-1,\nonumber\\
	& &\boldsymbol{u}=\boldsymbol{\chi},\quad\boldsymbol{v}= (\boldsymbol{\chi} \cdot \boldsymbol{k}) \boldsymbol{k}+\left(\mu_0\epsilon_0\omega^2-\boldsymbol{k}^2\right) \boldsymbol{\chi}.
\end{eqnarray}

%\begin{equation}\label{matriz_geral}
%\begin{aligned}
%	M_{i j} & =\left[\mu_0\epsilon_0\omega^2-\boldsymbol{k}^2+(\boldsymbol{\chi} \cdot \boldsymbol{k})^2\right] \delta_{i j}- \chi_i\left[(\boldsymbol{\chi} \cdot \boldsymbol{k}) k_j+\left(\omega^2-\boldsymbol{k}^2\right) \chi_j\right].
%\end{aligned}
%\end{equation}

The equation (\ref{eq_matricial}) allows non-trivial solutions if the matrix $M$ is singular, i.e., if $\operatorname{det} M=0$:
\begin{equation}
	\det M = \alpha^2\left(\alpha+\beta\boldsymbol{u}\cdot\boldsymbol{v}\right)=0. 
\end{equation}
The dispersion relations are derived from the secular equation of the matrix, initially yielding the conventional frequency $\omega(\boldsymbol{k}) = c |\boldsymbol{k}|$ and one modified solution given by
\begin{eqnarray}\label{RD1}
	\omega(\boldsymbol{k}) &=& c |\boldsymbol{k}|\sqrt{1-(\boldsymbol{\chi}\cdot\hat{\boldsymbol{k}})^2}.
\end{eqnarray}
The analysis of the asymptotic behavior of the modified expression allows for verification of its consistency with standard electrodynamics results. Taking the limit $|\boldsymbol{\chi}|\to 0$, we recover the Maxwell electrodynamic results, as expected. On the other hand, using the De Broglie correspondence, $\omega \leftrightarrow E / \hbar$ e $\boldsymbol{k} \leftrightarrow \boldsymbol{p} / \hbar$, it is possible to rewrite the equation as a function of energy and linear momentum. This yields the following dispersion relation in the energy-momentum space
\begin{equation}\label{RD2}
	E=c|\boldsymbol{p}| \sqrt{1-(\boldsymbol{\chi} \cdot \hat{\boldsymbol{p}})^2}.
\end{equation}

The Compton effect is a scattering process wherein the dispersion relations govern the energy and momentum exchange between the incident photon and the target particle. Eq. (\ref{RD2}) can be applied to study the increase in the photon's wavelength after being scattered by the electron, taken as the target, for the non-trivial LV parameter. From now on, we consider that $\hbar=c=1$ and $4 \pi \epsilon_0=1$. The initial state of the photon could be described by a wavelength $\lambda=1 /|\boldsymbol{p}|$, with energy $\boldsymbol{E}$, where the relation between $\boldsymbol{E}$ and momentum $\boldsymbol{p}$ must obey the results given by eq. (\ref{RD2}). We have the physical scenario in which one photon propagates in a spacetime characterized by this anisotropic background vector, and collides with an initially stationary electron. Before the interaction, we assume that the background vector exclusively modifies the photon's dispersion relation. The collision causes the electron to recoil with a certain energy and momentum. After scattering, the outgoing electron may couple to the background vector, and consequently, its dispersion relation no longer corresponds to that of a free particle. However, the present analysis focuses on the photon's wavelength shift and scattering angle; therefore, the effect of the background vector on the dynamics of the scattered electron is not addressed.

Following the collision, the photon's trajectory is deflected by a certain angle $\theta_{c,}$ with a wavelenght $\lambda^{\prime}=$ $1 /\left|\boldsymbol{p}^{\prime}\right|$ and energy $\boldsymbol{E}^{\prime}$. From energy and linear momentum conservation, the variation in the photon's wavelength after the collision process is given by
\begin{equation}\label{Cp_c}
	\lambda^{\prime}-\lambda=2 \lambda_e \sin ^2\left(\frac{\theta_c}{2}\right)+\frac{|\boldsymbol{\chi\cdot\hat{\boldsymbol{k}}}|^2}{2}\left[\lambda^{\prime}-\lambda+\frac{\lambda_e}{2} \frac{\left(\lambda^{\prime}-\lambda\right)^2}{\lambda^{\prime} \lambda}\right],
\end{equation}
where $\lambda_e=m_e^{-1}=2(\mathrm{MeV})^{-1}$ is the electron Compton wavelength and $|\boldsymbol{\chi}\cdot\hat{\boldsymbol{k}}|<1$. 

%If we assume $\lambda \gg \lambda_e$ on the equation (\ref{Cp_c}), the real and positive solution leads us for the wavelenght variation showed below

%\begin{equation}
%	\Delta \lambda \simeq 2 \lambda_e \sin ^2\left(\frac{\theta_c}{2}\right)\left[1-\frac{|\boldsymbol{\chi} \cdot \hat{\boldsymbol{p}}|^2}{2}\right]^{-1}.
%\end{equation}

We can analyze the limiting case of eq. (\ref{Cp_c}). If we rewrite this equation as follows:
\begin{equation}\label{CP1}
    (\Delta\lambda)^{2}\Big[4\lambda -2\lambda|\vec{\chi} \cdot \hat{k}|-\lambda_{e}|\vec{\chi} \cdot \hat{k}|^{2}\Big]+(\Delta\lambda)\Big[4\lambda^{2}-8\lambda_{e}\lambda\sin^{2}\Big(\frac{\theta_{c}}{2}\Big)-2\lambda^{2}|\vec{\chi} \cdot \hat{k}|^{2} \Big]-8\lambda_{e}\lambda^{2}\sin^{2}\Big(\frac{\theta_{c}}{2}\Big)=0,
\end{equation}
where $\Delta\lambda=\lambda'-\lambda$, as we can see, we have a quadratic equation in $\Delta\lambda$. After alengthy calculation, we found the result
\begin{eqnarray}\label{CP2}
\Delta\lambda&=&\Big[32\lambda^{2} \lambda_{e}\sin^{2}\Big(\frac{\theta_{c}}{2}\Big)\Big(-\lambda_{e}|\vec{\chi}\cdot\hat{k}|^{2}-2\lambda|\vec{\chi}\cdot\hat{k}|^{2}+4\lambda\Big)+\nonumber\\&+&
\Big(-2\lambda|\vec{\chi}\cdot\hat{k}|^{2}-8\lambda\lambda_{e}\sin^{2}\Big(\frac{\theta_{c}}{2}\Big)+4\lambda^{2}\Big)^{2}}\Big]^{1/2}
[2\Big(\lambda_{e}|\vec{\chi}\cdot \hat{k}|^{2}+2\lambda|\vec{\chi}\cdot \hat{k}|^{2}-4\lambda\Big)]^{-1}{\nonumber 
\\&& 
-\frac{\Big(+2\lambda|\vec{\chi}\cdot \hat{k}|^{2}+8\lambda_{e}\lambda\sin^{2}\Big(\frac{\theta_{c}}{2}\Big)-4\lambda^{2}\Big)}{2\Big(\lambda_{e}|\vec{\chi}\cdot \hat{k}|^{2}+2\lambda|\vec{\chi}\cdot \hat{k}|^{2}-4\lambda\Big)}
\end{eqnarray}
and
\begin{eqnarray}\label{CP3}
\Delta\lambda^{'}&=&-\Big[32\lambda^{2} \lambda_{e}\sin^{2}\Big(\frac{\theta_{c}}{2}\Big)\Big(-\lambda_{e}|\vec{\chi}\cdot\hat{k}|^{2}-2\lambda|\vec{\chi}\cdot\hat{k}|^{2}+4\lambda\Big)+\nonumber\\&+&
\Big(-2\lambda|\vec{\chi}\cdot\hat{k}|^{2}-8\lambda\lambda_{e}\sin^{2}\Big(\frac{\theta_{c}}{2}\Big)+4\lambda^{2}\Big)^{2}\Big]^{1/2}
%\times\nonumber\\&\times&
\Big[2\Big(\lambda_{e}|\vec{\chi}\cdot \hat{k}|^{2}+2\lambda|\vec{\chi}\cdot \hat{k}|^{2}-4\lambda\Big)\Big]^{-1}\nonumber \\&& 
-\frac{\Big(+2\lambda|\vec{\chi}\cdot \hat{k}|^{2}+8\lambda_{e}\lambda\sin^{2}\Big(\frac{\theta_{c}}{2}\Big)-4\lambda^{2}\Big)}{2\Big(\lambda_{e}|\vec{\chi}\cdot \hat{k}|^{2}+2\lambda|\vec{\chi}\cdot \hat{k}|^{2}-4\lambda\Big)}.
\end{eqnarray}
A comment is in order. We need to consider just the positive square root (\ref{CP2}) as a physically acceptable solution since scattered particles must lose part of their energy. We can check the result taking the limit $|\boldsymbol{\chi}|\to 0$, where we found the following result
\begin{equation}\label{CP4} \Delta\lambda=\frac{(1/2)\sqrt{+128\lambda_{e}\lambda^{3}\sin^{2}\Big(\frac{\theta_{c}}{2}\Big)+\Big(4\lambda^{2}-8\lambda_{e}\lambda\sin^{2}\Big(\frac{\theta_{c}}{2}\Big)\Big)^{2}}+4\lambda_{e}\lambda\sin^{2}\Big(\frac{\theta_{c}}{2}\Big)-2\lambda^{2}}{4\lambda}
\end{equation}
and
\begin{equation}\label{CP5} 
\Delta\lambda^{'}=\frac{-(1/2)\sqrt{+128\lambda_{e}\lambda^{3}\sin^{2}\Big(\frac{\theta_{c}}{2}\Big)+\Big(4\lambda^{2}-8\lambda_{e}\lambda\sin^{2}\Big(\frac{\theta_{c}}{2}\Big)\Big)^{2}}+4\lambda_{e}\lambda\sin^{2}\Big(\frac{\theta_{c}}{2}\Big)-2\lambda^{2}}{4\lambda},
\end{equation}
that, after some simplifications, and considering just the equation (\ref{CP4}) as a physically acceptable solution, we recover the standard wavelength shift of the photon in the Compton effect:
\begin{equation}
    \Delta\lambda=2\lambda_{e}\sin^{2}\Big(\frac{\theta_{c}}{2}\Big)
\end{equation}
as expected\footnote{Eq. (\ref{CP5}), corresponding to another solution, gives the result $\Delta\lambda^{'}=-\lambda$, which is not physically acceptable solution.}.

\section{Concluding Remarks}
We considered the CPT-even LV electrodynamics. In this theory, we described, for the first time in LV theories, the Compton effect and calculated the photon wavelength shift, which turned out to be corrected by an additive term quadratic in the LV parameter. While this effect is very small, it is natural to expect that its experimental study could be used to impose stringent bounds on the LV parameter contributing to the results presented in \cite{Kostelecky:2008ts}. In principle, an analogous study can be performed as well for other LV extensions of QED, which we plan to study in a forthcoming paper.

{\bf Acknowledgments.}  The work of A. Yu.\ P. has been partially supported by the CNPq project No. 303777/2023-0.  The work by J. C. C. F. has been partially supported by FAPEMIG.

\bibliographystyle{ieeetr}
\bibliography{referencias}

\end{document}